\documentclass[fleqn,usenatbib,useAMS]{mnras}

\usepackage{graphicx}   
\usepackage{amsmath}    
\usepackage{amssymb}    
\usepackage{multicol}        
\usepackage{bm}         
\usepackage{pdflscape}  
\usepackage{xcolor}
\usepackage{bm}

\newcommand{\hmpc}{h^{-1}{\rm ~Mpc}}

\usepackage[T1]{fontenc}
\usepackage{ae,aecompl}

\usepackage{txfonts}

\title[The clustering of voids]{The sparkling Universe: 
Clustering of voids and void clumps}

\author[Lares et al.]{\parbox[t]{\textwidth}{\vspace{-1cm}
        Marcelo Lares\thanks{e-mail: marcelo.lares@unc.edu.ar}, Andr\'es N. Ruiz, Heliana E. Luparello, Laura Ceccarelli, 
        Diego Garcia Lambas \& Dante J. Paz}\vspace{0.2cm}\\
Instituto de Astronom\'{\i}a Te\'orica y Experimental (CONICET-UNC) and 
Observatorio Astron\'omico (UNC), Laprida 854, X5000BGR, C\'ordoba, Argentina.
}

\begin{document}

\date{Accepted XXX. Received XXX; in original form XXX}
\pubyear{2015}
\label{firstpage}    
\pagerange{\pageref{firstpage}--\pageref{lastpage}} \pubyear{XXXX}

 \maketitle

\begin{abstract}
%
We analyse the clustering of cosmic voids using a numerical simulation
and the main galaxy sample from the Sloan Digital Sky Survey.
We take into account the classification of voids into two types that
resemble different evolutionary modes: those with a rising
integrated density profile (void--in--void mode, or R--type) and voids
with shells (void--in--cloud mode, or S--type).  
The results show that voids of the same type have stronger clustering
than the full sample.
We use the correlation analysis to define void clumps, associations
with at least two voids separated by a distance of at most the mean
void separation.
In order to study the spatial configuration of void clumps, we compute
the minimal spanning tree and analyse their multiplicity, maximum
length and elongation parameter. 
We further study the dynamics of the smaller sphere that encloses all
the voids in each clump.         
Although the global densities of void clumps are different according
to their member--void types, the bulk motions of these spheres are
remarkably lower than those of randomly placed spheres with the same
radii distribution.
In addition, the coherence of pairwise void motions does not strongly
depend on whether voids belong to the same clump. 
Void clumps are useful to analyse the large--scale flows around voids,
since voids embedded in large underdense regions are mostly in the
void--in--void regime, were the expansion of the larger region
produces the separation of voids.
Similarly, voids around overdense regions form clumps that are in
collapse, as reflected in the relative velocities of
voids that are mostly approaching.
%
\end{abstract}


\section{Motivation}
\label{S_intro}

The structure of the mass distribution in the Universe at large scales
can be described as a network with a typically filamentary structure,
which intersect forming even denser clumps, the preferred places where
galaxy clusters are formed.
The origin and evolution of this arrangement can be understood in the
framework given by the cosmological models, and supported by
observational evidence \citep[e.g.][]{liddle_introduction_2003,
dodelson_modern_2003} which depict nearly the same picture of
hierarchical structure formation \citep{padmanabhan_structure_1993,
peebles_priciples_1993}.
Among the currently discussed scenarios in the literature, the
concordance $\Lambda$CDM is the preferred model at the present, given
its ability to predict a large variety of observed phenomena
\citep{weinberg_gravitation_1972}.
As the Universe evolves, mass is accreted onto the densest
concentrations, giving rise to large empty regions in the distant
future \citep{dunner_limits_2006, arayamelo_future_2009,
pearson_extent_2015}.
A different picture of the same process is obtained by looking at the
initial low density fluctuations that become increasingly emptier,
larger and rounder, as mass flows towards dense regions. 
These two pictures are complementary and manifest in the large--scale
distribution of matter \citep{einasto_structure_1986,
einasto_supercluster_1997}, producing a filamentary--void network
\citep[e.g.][]{matsuda_topology_1984, way_structure_2011,
 icke_galaxy_1991, leclerq_cosmic_2015}.
According to this scenario, not only large--scale flows of mass play a
key role in shaping the largest structures, but also they are
intimately connected to the mass distribution itself.
This reflects the reciprocal action between the source of gravity and
the forces it produces, as described by the field equations of general
relativity.
The scales at which this action can be detected are considerable large
\citep{watkins_consistently_2009, feldman_cosmic_2010,
nusser_bulk_2011, turnbull_cosmic_2012}.
According to this model, large--scale flows of mass play a key role in
forming the largest structures and shaping the cosmic web, and its
effects can be detected up to considerable large scales
\citep{frisch_evolution_1995, watkins_consistently_2009, feldman_cosmic_2010,
nusser_bulk_2011, turnbull_cosmic_2012}.

In \citet{lambas_sparkling_2016} we reported the motions of cosmic
voids as a whole, which also show a strong coherence pattern
associated to the void velocity field up to large cosmological scales,
both in simulations and observations.          
This effect strongly depends on the type of void considered, with a
void--in--void and void--in--cloud classification scheme proposed by
\citet{sheth_hierarchy_2004} that distinguish the internal dynamical
behaviour \citep{paz_clues_2013} according to their environment
\citep{ceccarelli_clues_2013}.
The coherence pattern in the relative velocities is twofold once voids
are divided according to this classification: void coherent bulk
velocities define a bimodal dynamical population of mutually
attracting (for shell like voids) or receding (for voids embedded on
large-scale underdensities) systems.
We argue that these global motions contribute to imprint large scale
cosmic flows that will shape the formation of structures in the
distant future.

In what follows, we recap the properties of the catalogues of voids
and study the clustering of voids by computing the 2--point
autocorrelation function of voids (Sec. \ref{S_void_clustering}).
Given the significant correlations between voids, a percolation
algorithm is suitable to identify conspicuous groups of voids.
We then analyse these groups, which we call ``void clumps'', through
their geometrical and dynamical properties (Sec. \ref{S_clumps}).
The dynamical behaviour of void clumps is not equivalent to other
regions of the same volume centred in random locations (Sec.
\ref{S_envir}).  
Thus, we explore the global dynamics of the regions occupied by the
clumps and their internal motions of mass as a function of the type of
voids which compose the clumps.
Finally, in the Sec. \ref{S_discuss}, we discuss the results in the
context of the hierarchical structure formation scenario.

\begin{figure}
\includegraphics[width=0.5\textwidth]{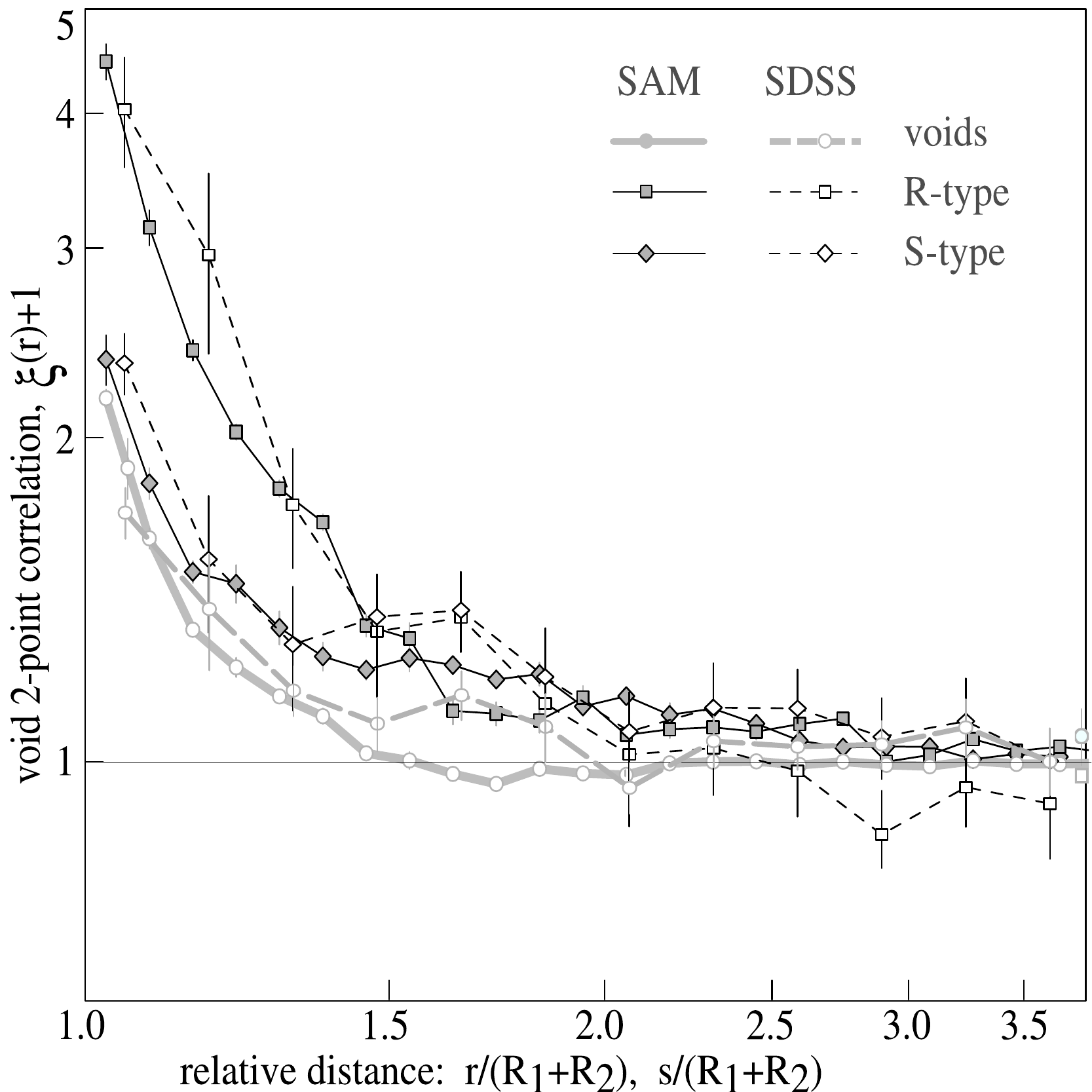}
\caption{Autocorrelation functions of voids, using the full sample
   (grey lines), S--type voids (diamonds) and R--type voids
   (squares), for simulation (solid lines, filled symbols) and SDSS data (dashed lines, empty symbols).
   Distances are in units of $\hmpc$, in real space for the simulation and in redshift space for SDSS data.
   Error bars correspond to uncertainties computed by means of Jackknife resampling.} 
\label{F_1} 
\end{figure}
 
\begin{figure*}
\includegraphics[width=0.95\textwidth]{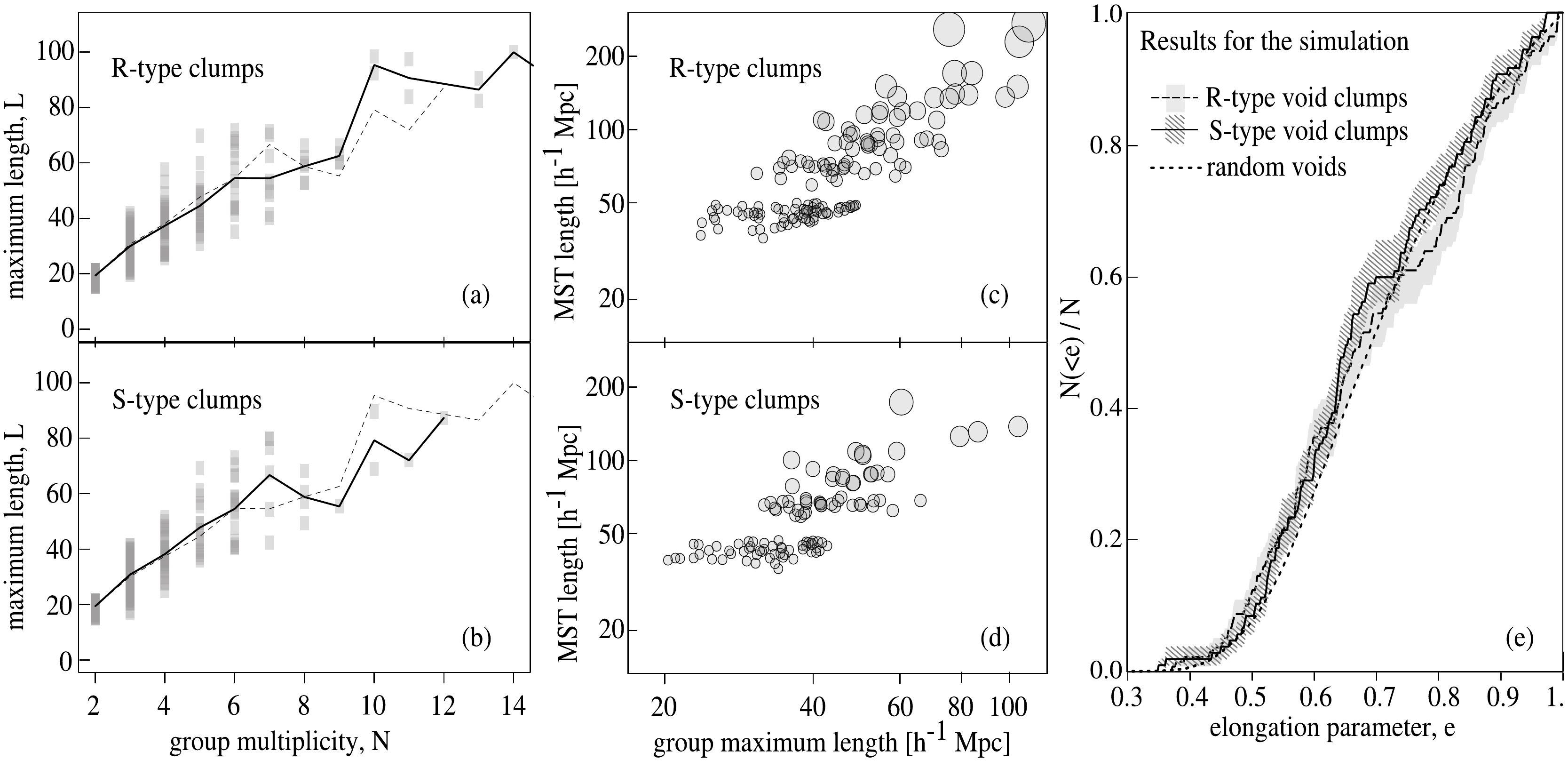}
\caption{
   {\it Left:} Relation between the maximum separation of voids within
   a clump and clump multiplicity, for R--type void clumps (a) and
   S--type void clumps (b).  Each grey square corresponds to one
   group.  The means of maximum lengths for each multiplicity are also
   shown in solid lines, and repeated on both panels for comparison
   (dashed lines).
   {\it Centre:} Relation between the length of the minimal spanning
   tree (MST) and the maximum length for R--type void clumps (c) and
   S--type void clumps (d).  The circle size is proportional to the
   clump multiplicity. 
   {\it Right:}  Empirical cumulative distributions of the elongation
   parameter, for R--type (dark solid lines) and S--type void clumps
   (dark dashed lines).  Dotted line corresponds to the distribution
   for groups of random points with the same multiplicities than void
clumps.}
\label{F_2} 
\end{figure*}

\section{Data}
\label{S_data}

In this section we describe the galaxy catalogues, both observed and simulated,
and the void identification algorithm used in this work. 

\subsection{Galaxy catalogues}
\label{SS_catalogues}

We use the Main Galaxy Sample of the Sloan Digital Sky
Survey Data Release 7 \citep[SDSS-DR7,][]{abazajian_seventh_2009}. 
This sample counts with nearly a million of galaxies with
spectroscopic measures, redshifts up to $z \le 0.3$ and an upper
apparent magnitude in the $r$-band of 17.77.
For this Main Sample, we select galaxies with a limiting redshift
$z=0.08$ and a maximum absolute magnitude in the $r$-band of
$M_r-5\log_{10}(h) = -19.1$. 
This guaranties a volume complete galaxy sample at that redshift.
The SDSS velocity field we consider, corresponds to the peculiar velocity
field derived by
\citet{wang_reconstructing_2009,wang_reconstructing_2012}, where the
authors employed the linear theory relation between mass
overdensities and peculiar velocities to reconstruct the 3D velocity
field of galaxies.
A detailed analysis of the effects of the linearised velocity field in
estimating void bulk velocities in the observational sample can be
seen in the Appendix of \citet{ceccarelli_sparkling_2016}.

We also use the semi-analytical galaxies presented by
\citet{guo_sam_2011}, which were constructed by applying the Munich
semi-analytic model (SAM) of galaxy formation to the dark matter only
Millennium Simulation \citep[MS, ][]{springel_ms_2005}.
The MS counts $2140^3$ dark matter particles evolved from $z=127$
to $z=0$ in a cubic comoving volume of $(500\hmpc)^3$.
The cosmological parameters used in the MS correspond to a
$\Lambda$CDM flat cosmology with $\Omega_{\rm m} = 0.25$,
$\Omega_\Lambda = 0.75$, $\Omega_{\rm b} = 0.045$, $\sigma_{\rm 8} =
0.9$, $h=0.73$ and $n=1.0$. This parameters are consistent with the
WMAP1 results \citep{spergel_wmap1_2003}.
The galaxy catalogue of \citet{guo_sam_2011} is public available at
the Millennium
Database\footnote{http://gavo.mpa-garching.mpg.de/Millennium}.

In order to make a fair comparison between observations and the
simulated data, we analyse SAM galaxy samples with the same
number density than the observed galaxy distribution 
\citep[see e.g.][]{contreras_density_2013,contreras_density_2015}. 
Using the same magnitude cut in the SAM galaxies than in SDSS does not
guaranties the same number density, because of the differences between 
their luminosity functions.  
Instead, we select all SAM galaxies brighter than $M_r-5\log_{10}(h) = -19.7$, 
which guarantees the volume density needed.


\subsection{Void identification}
\label{SS_void_id}

The identification of voids was performed following the procedures described
in \citet{ruiz_clues_2015}, which is a modified version of the
algorithms presented in \cite{padilla_spatial_2005} and
\cite{ceccarelli_voids_2006}.

The identification starts using the galaxy catalogues as tracers of
the density field and constructing a contrast density field estimation
using a Voronoi tessellation, selecting as void candidates all the
underdense cells with a density contrast bellow $-0.8$.
We identify voids both in a numerical simulation and in the SDSS data.
Centred in these underdense cells, we compute the integrated density
contrast $\Delta(r)$ at increasing values of radius $r$, and select as
void candidates the largest spheres which satisfy the condition
$\Delta(R_{\rm void})<-0.9$, with $R_{\rm void}$ as the void radius.
In order to recentre each candidate, the centre position is randomly
shifted and the procedure described previously is repeated in this new
centre, allowing the sphere to grow in size. 
Finally, a void of radius $R_{\rm void}$ is selected as the largest
sphere satisfying the underdense condition that does not overlap any
other underdense sphere.
In the case of SDSS data, we restrict our void definition
to spherical regions with a fixed global density, excluding
those spheres that are not completely within the survey mask. 
The final catalogues comprise 252 voids for the SDSS sample and 4015
voids for the full Millennium box, both with similar radii
distributions in the range 6--24 $\hmpc$.
The larger number of voids in the Millennium box simulation is
consistent with the volume difference with the SDSS data.



\begin{figure}
\centering \hfill
\includegraphics[width=0.47\textwidth]{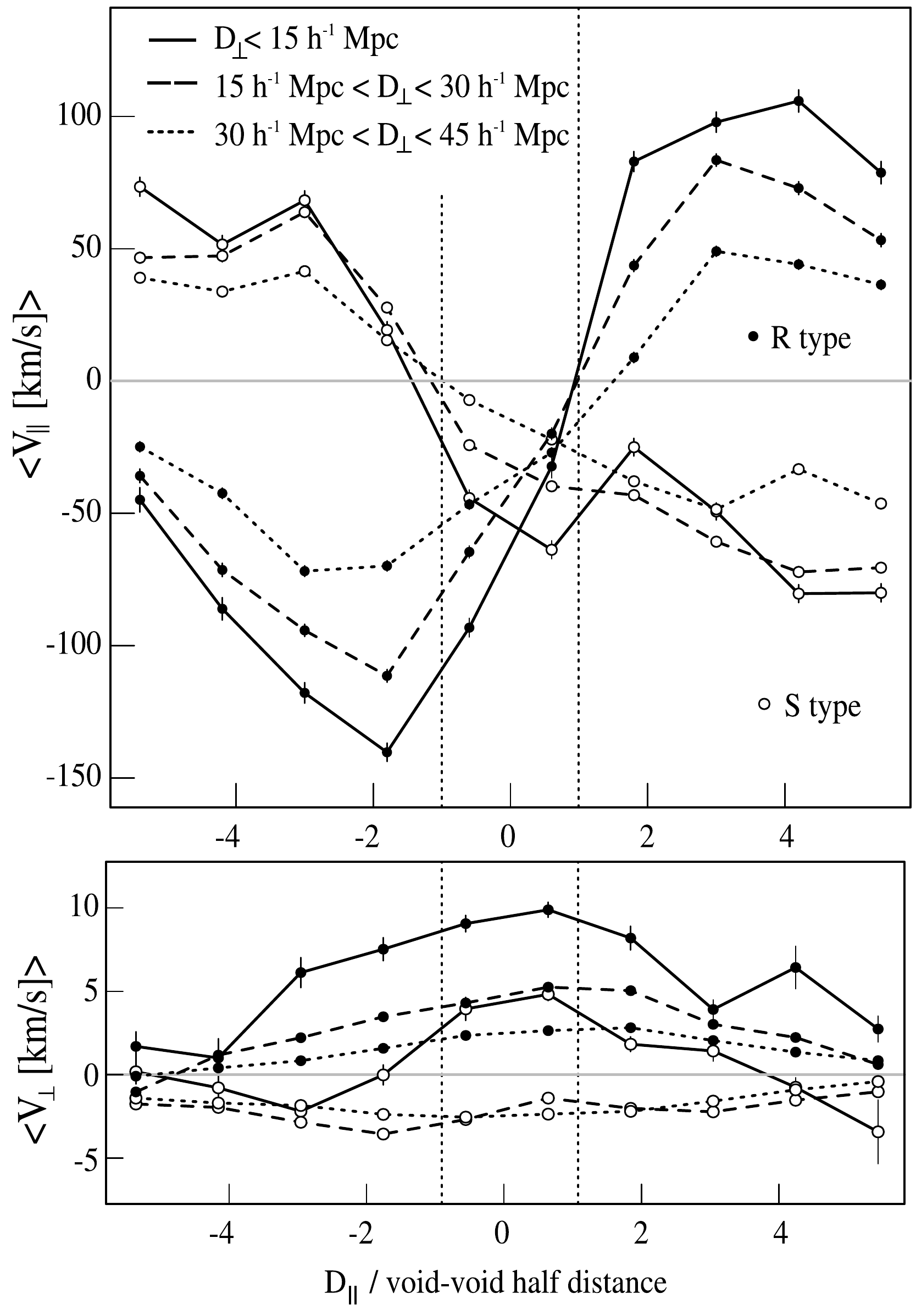}
\caption{Mean velocity components of SAM galaxies in the directions
parallel (upper panel) and perpendicular (bottom panel) to the
direction between the two void centres, in clumps formed by two
R--type voids (filled circles) or two S--type voids (empty circles).
The different curves (see figure key) correspond to galaxies within 
a perpendicular separation D$_{\perp}$ range to the axis along the 
two void members (i.e. cylindrical shells).
Vertical dotted lines indicate the position of each void in the pair.
}
\label{F_3} 
\end{figure}

\section{Void clustering}
\label{S_void_clustering}

As mentioned in Sec. \ref{S_intro}, the gravitational growth of the
large scale structure can be described by means of
two complementary scenarios, 
the accretion of mass onto massive objects and
the expansion of underdense regions. 
In this context, we analyse here the spatial distribution of voids 
focusing on its relation to large scale flows.
Previous results
have shown a coherent velocity field of voids
\citep{lambas_sparkling_2016}, an effect that can be explained in
terms of the large--scale surrounding distribution of mass
\citep{ceccarelli_sparkling_2016}.
Therefore, it is natural to expect a relation between the clustering of
mass, large--scale velocity flows, and the clustering of underdense regions.

It must be taken into account that the void definition used in our
study comprises regions that contain at most the 10 per cent of the
mean density of the Universe. 
The clustering of voids, as analysed here, is thus a manifestation of
the locations of almost empty regions, and has a valuable information
from a cosmological perspective since it will eventually allow to
study the volumes, shapes, percolation and density profiles of the
largest underdense regions at a fixed degree of underdensity.

The correlation function, $\xi(r)$, measures the probability excess of
finding a pair of objects at a given relative distance $r$ with
respect to a random distribution.
This tool has been extensively used to quantify the clustering of
galaxies \citep{einasto_supercluster_1997, kerscher_comparisson_2000} and is a key observable to
distinguish cosmological models and test the structure formation
scenarios \citep{matsubara_correlation_2004}.
In a pioneer work, \citet{padilla_spatial_2005} examined the void-void
correlation function in a numerical simulation.
They compared the clustering of haloes and galaxies finding larger
voids to be strongly clustered.
However the correlation amplitude is not statistically significant due
to the small simulation box.
Similarly, \citet{clampitt_clustering_2016}, using the autocorrelation
function of SDSS and simulated voids, also find a stronger signal for
the larger voids.
There are other algorithms to identify voids that are not restricted
to the spherical condition.
A usually adopted algorithm is the ZOBOV finder
\citep{neyrinck_zobov_2008}, which makes
a Voronoi tessellation of the space to estimate de density field.
Voids are identified through a watershed algorithm, as regions around
local minima limited by ridges in the density field that separate different
minima.
The clustering of ZOBOV voids has been analysed in simulations
\citep{hamaus_cosmology_2014, chan_large_2014, hamaus_testing_2014}.
%
%
%
%
%
%
\citet{hamaus_cosmology_2014} analyse the void bias 
and describe two
different populations: small voids, which are overcompensated by the mass in
the surrounding regions, and large, undercompensated voids.
While small voids have large bias with respect to the dark matter
distribution, larger voids are preferentially anti--correlated.
The authors argue that the high bias in small voids is due to an
overcompensation of void shells around small voids, that are typically
voids--in--clouds.
This resembles the S-type classification we use in our work, although we find
that while most of the large voids are of R--type,  small voids can be of
either type in nearly equal number \citep[see][]{ceccarelli_clues_2013}. 
In spite of these differences, this classification is also based on the spirit
of the void--in--cloud (overcompensated) and void--in--void (undercompensated)
types and results in different dynamical evolution and different bias
parameters.
However, \citet{hamaus_cosmology_2014}  show that the distribution of voids is
dominated by a
Poisson noise for small voids, and with smaller power for large voids given by
the exclusion effects.
\citet{hamaus_testing_2014} expand these ideas and use the void clustering
statistic in a cosmological simulation to probe the cosmic expansion history of
the universe.
These features could be explained by an excursion set formalism \citep{chan_large_2014}.
\citet{zhao_dive_2016} use a different void finder based on a Delaunay
triangulation of a set of tracers (DIVE) and identify two different
populations of voids characterised by their radii, which resemble the
void--in--void and void--in--cloud regimes.
The authors compute the power spectrum of DIVE voids and find that,
on large scales, 
large voids show a low bias while small voids are strongly
biased, in agreement with previous results.
The clustering of voids has also been analysed by
\citet{liang_measuring_2016}, where the authors find a conspicuous
signal at nearly $100\hmpc$ which indicates the presence of the Baryon
Acoustic Oscillations of mass in the early Universe.
The definition of voids in their study does not account for the
so called exclusion effect, i.e., all overlapping underdense spheres
are considered.
The authors also find a scale dependent bias for different samples
of voids depending on void radius, with the larger voids showing the
strongest signal.
This favours a larger clustering signal for voids in the
void--in--void regime.
This result was confirmed later by \citet{kitaura_signatures_2016}, who 
analyse overlapping density troughs of the density field and argue
that the detection of baryonic acoustic oscillations
is not significant for the classical
definition of disjoint voids.

In this work, we use the correlation function of the distribution of
distances between the centres of pairs of voids to measure the degree
of clustering of voids.
In order to compute this function, we counted pairs in bins of
relative distance, that is the comoving separation between void
centres divided by the sum of its radii.  Here, a separation of
$r/(R_1+R_2)=1$ means that the pair of voids are in contact, with a
separation between centres equal to the sum of their radii.
The normalization of the correlation function to the sum of the void
radii is convenient since the signal would have a mixture of different
contributions from small and large voids in natural units of distance.
To construct these functions, we generated mock catalogues in the same
box than that of the simulation, using a simple sequential inhibition
algorithm to reproduce the exclusion effect produced by the finite
size of the voids.
This is important since the scales of interest, where the correlation
signal is observed, is comparable to the size of the voids.
The results, applied to our sample of voids, are shown in Fig.
\ref{F_1} for several samples of voids in the simulation and in the
SDSS galaxy catalogue.
We compute $\xi(r)$ (simulation) and  $\xi(s)$ (observations) 
for relative distances larger than the unity, due
to the exclusion imposed in the identification algorithm (see Sec.
\ref{SS_void_id}).
The dashed (solid) grey line, show the results for the autocorrelation
of voids identified on the SDSS (SAM catalogue).
Beyond roughly a relative distance of 2, the distribution of pairs of
voids is consistent with a Poisson distribution.
On the other hand, there is a range of distances with a significant
excess of void pairs, for a typical void size of $10$-$12\hmpc$ this
corresponds to scales between $20$ and $45\hmpc$.
This scale is larger than that of the void shells reported in
\citet{paz_clues_2013}.

In order to explore the role of the environment on the correlation, we
separated voids according to the criteria presented in
\citet{ceccarelli_clues_2013}, that in turn follows the ideas proposed
by \citet{sheth_hierarchy_2004}.
Voids that have a steep integrated density profile resembling a
shell--like surrounding structure are classified as S--type, while
voids with a gently rising profile are classified as R--type.
This classification was used in previous works
\citep{ceccarelli_clues_2013,  paz_clues_2013, ruiz_clues_2015,
lambas_sparkling_2016,ceccarelli_sparkling_2016}, where it was proved
effective at separating two distinct populations of void environments.
In Fig. \ref{F_1} we show the autocorrelation of R--type (squares) and
S--type (diamonds) voids for SDSS (dashed lines) and SAM data
(solid lines).
Remarkably, the autocorrelation of R--type voids is significantly
higher than that of the general population 
within $\mathrm r \gtrsim$ 2 (R$_1$+R$_2$),
indicating that this type
of voids are preferentially clustered.
At distances larger than  $\mathrm r \gtrsim$ 2 (R$_1$+R$_2$), there is
no evidence of a significant difference between the sample populations.  This
is related to the relatively small volume of the survey data, which gives large
uncertainties as indicated by the error bars of the lines corresponding to SDSS
galaxies.
Similarly, but with lower significance, S--type voids tend
to be more clustered than the general void population.

We compute the statistical uncertainties of the correlation on each
distance bin using Jackknife or leave--one--out resampling.
For a sample of voids with size $N_v$, we use the ''natural''
estimator \citep{kerscher_comparisson_2000} of the correlation
function, which estimates the correlation function $\xi(i)$ in terms
of the number of void pairs ($DD(i)$) and the number of random void
pairs ($RR(i)$) at the i--th distance bin that spans from $r_i$ to
$r_{i+1}$.
The central tendency measure of $\xi(i)$ is given by the Jackknife
average \citep{efron_jackknife_1987, lupton_statistics_1993}:

\begin{equation}
   \bar{\xi}_J(i) = \frac{1}{N_v - 1} \sum_{j=1}^{N_v} \xi_{[j]}(i),
\end{equation}

\noindent where $\xi_{[j]}(i)$ is the j-th Jackknife realization,
i.e., the value of $\xi$ computed by leaving the j-th element element
out.  
The number of Jackknife realizations is equal to the number of voids
in the sample.
For a Jackknife realization the data-data pair counts turn to be
$DD_{[j]}$.  
The random-random pairs $<RR(i)>$ are not affected by the Jackknife
step since we use the average of a large number of void sample mocks.
Then, the correlation function estimation for the j-th Jackknife
realization is given by:

\begin{equation}
   \xi_{[j]}(i) = \frac{DD_{[j]}(i)}{<RR(i)>} \, \frac{N_v}{N_v-1} - 1.
\end{equation}

\noindent 

The compensation factor $N_v/(N_v-1)$ compensates for the different
number of centre voids in the centre-tracer scheme of pair
computations.
Since it is an autocorrelation and the random samples are generated
with the same number of voids for each void sample, the number of
tracer is the same and no other correction is required.
Finally, the uncertainty is estimated by:

\begin{equation}
   \hat{\sigma}_J^2(i) = \frac{N_v-1}{N_v} \sum_{j=1}^{N_v} (
   \bar{\xi}_{[j]}(i) - \bar{\xi}_{J}(i) )^2
\end{equation}


\begin{figure}
\includegraphics[width=0.45\textwidth]{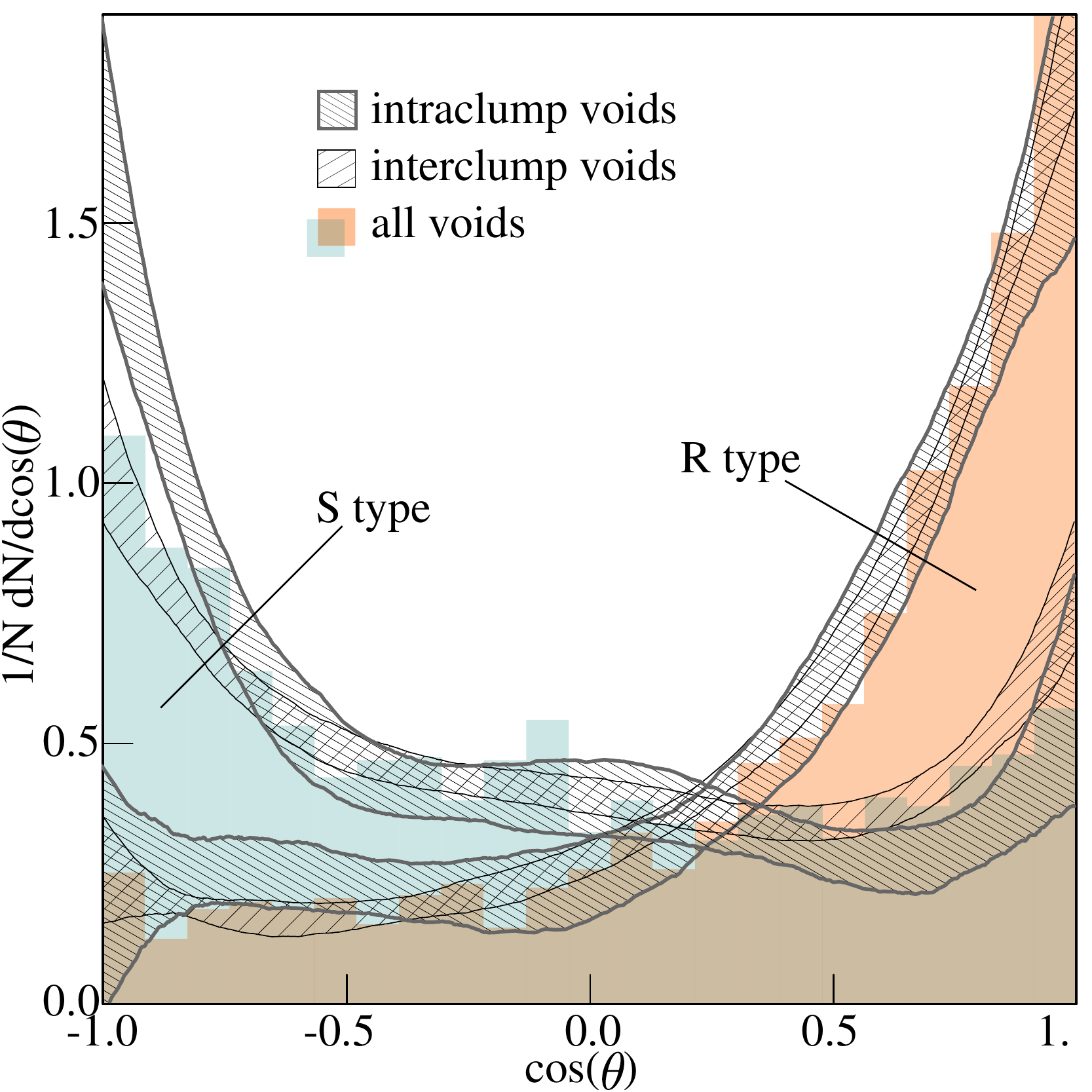}
\caption{Number density of void pairs as a function of $\cos(\theta)$
   for all voids in the 
   the simulation.  We show separately the estimations of the distributions of
   $\cos(\theta)$, 
   with histograms for all R-- and S--type voids and with smoothed 
   curves for voids within the same clump or in different clumps (see figure key).  
   The shaded regions are estimations of the
   distributions after a dimensionality reduction with a Fourier expansion and
   represent 1-$\sigma$ bootstrap uncertainties. The results for R--type voids and
   S--type voids are shown separately, where S--type distributions favour negative
   values of $\cos(\theta)$, corresponding to approaching pairs, and
   R--type voids are
   dominated by positive values, corresponding to receding pairs.
}
\label{F_4} 
\end{figure}

\section{Clumps of voids}
\label{S_clumps}

In this Section we present the definition and identification procedure
used to detect void clumps both in the simulation and SDSS data.
Also we present an analysis of its geometrical properties.

\subsection{Identification procedure}
\label{SS_clumps_id}

The detection of clustering in the void distribution (as shown in Sec.
\ref{S_void_clustering}) leaded us to apply a percolation algorithm in
order to isolate groups of voids, and obtain insight on the origin of
the higher correlation observed for the two void types.
We searched for groups of voids by implementing the
friend--of--friends algorithm.
This is a simple algorithm that links together all voids that have at
least a subset pair having centres closer than a given linking length,
hereafter $\ell$.
The numerical value of this length, however, depends on the nature of
the problem and must be determined.
For both samples (R-- and S--type voids) we start by considering the
mean inter-void separation $\ell_{MVS}$:
\begin{equation} \ell_{MVS}=\left(\frac{3}{4\pi n}\right)^{1/3}
\end{equation}
where $n$ is the number density of a given void sample.  This value
can be interpreted as the radius of a sphere placed at random
containing on average one void
centre.
For the R--type sample of voids we obtain a value
$\ell_{MVS}=21.64\hmpc$, whereas for S--type voids
$\ell_{MVS}=21.47\hmpc$ is obtained.
Then, we explored several values of the linking length parameter
$\ell$ as a variable fraction $f$ of $\ell_{MVS}$ (i.e.
$\ell=f\,\ell_{MVS}$). 
For each case we computed the multiplicity
function, i.e., the distribution of multiplicities of the resulting
groups.
By taking values of $f<1$, a larger abundance of S--type void clumps
are obtained in comparison to the R--type clumps for the whole
range of multiplicities.
On the other hand, for $f>1$ the opposite behaviour is observed,
namely,
multiplicities are larger for R--type void clumps. Therefore, we defined
R--type and S--type void clumps as friend--of--friends groups with a linked length
equal to $\ell_{MVS}$, which preserves similar multiplicity
distributions for both samples. 


\subsection{Properties of void clumps}
\label{SS_clumps_prop}

We study the geometrical properties of the void clumps by
estimating their size and shape.
Since the number of members of the groups is typically low, usual
methods to compute the shape tensor do not work \citep{paz_shapes_2006}.
Instead, we computed the minimal spanning tree \citep[hereafter MST,
][]{kruskal_mst_1956} of the centres of voids in each group, and
compared it to the maximum separation between any two members.
The MST is the graph of minimal length which connects all members in
the group.
If the  group is very elongated, the numerical values of both measures
are similar, otherwise, the length of the MST
is larger than the maximum length separation.

Panels (a) and (b) of Fig. \ref{F_2} show the maximum length, L,
defined as the maximum separation between any pair of void centres
that belong to the same void clump, as a function of group
multiplicity, for R--type and S--type void clumps respectively, 
These results are repeated in both panels (dashed lines) in order to
allow comparison among void types.
Each grey square corresponds to a clump.
Solid lines show the mean of the maximum lengths for each multiplicity
value.
As expected, more populated clumps tend to be larger, with no
significant difference between the two types of voids.

Another estimator of the size of a void clump is the length of the
MST, which is the sum of its edge longitudes.
In panels (c) and (d) of Fig. \ref{F_2} we show the MST length for
R--type and S--type void clumps, respectively, as a function of L.
The size of the filled circles is proportional to the multiplicity of
the void clump.
All triplets have an MST length of order 40$\hmpc$, although the
maximum length ranges from 20$\hmpc$ to 40$\hmpc$. 
Similarly, more populated clumps have a larger range of values of the
maximum length than in the MST length, which is also verified for both
types of void clumps for a fixed multiplicity.
However, when comparing the R--type and S--type cases, there is a
slight excess of R--type clumps for maximum lengths over 50$\hmpc$ and
MST lengths over 100$\hmpc$.

The MST length can be compared to the maximum length to get an insight
on the shape of the clump.
According to this, an elongation parameter can be defined as the ratio
of the maximum separation length and the length of the MST for each
clump.
In panel (e) of Fig. \ref{F_2} we show the cumulative distribution of
this parameter.
The results for clumps of R--type voids are shown in dark solid lines,
and for clumps of S--type voids with dark dashed lines.
%
%
Errors represent 1-$\sigma$ uncertainties from a bootstrap resampling
estimation.
By applying suitable tests, we find no compelling
evidence that the observations and the random case are
different, independent of multiplicity.               


\subsection{Dynamics of galaxies in two-void clumps}
\label{SS_clumps_halos}

In this subsection, we analyse the dynamics of SAM galaxies in the
region surrounding clumps of N=2.
As mentioned in Sec. \ref{SS_catalogues}, we consider a limited
magnitude sample of galaxies in order to match the number density of
the SDSS volume limited sample. 
We define two velocity components: $\bm{V}_{\parallel}$, along the
line containing the two void centres; and $\bm{V}_{\perp}$, the
projection onto the plane normal to this line.
The sign of $\bm{V}_{\parallel}$ is defined positive in the direction
to the largest void (i.e. positive velocities are from left to right).
For each clump we also define the system centre as the position in the
middle of two void members.
For each galaxy we compute the two cylindrical components of its
separation vector to the system centre, one component along the line
of the two void members, $\bm{D}_{\parallel}$, and the other,
$\bm{D}_{\perp}$, as a cylindrical radial component on the plane
normal to this line.
We stack all clumps with N=2 for each type, and consider the
normalized projected distance, $d$, along the direction of the two
voids: $d=|\bm{D}_{\parallel}$|~/~void-void half distance.
With this definition, the smallest void is at $d=$-1 and the largest
at $d=$1.
In Fig. \ref{F_3} we plot the mean $\bm{V}_{\parallel}$ (upper panel)
and $\bm{V}_{\perp}$ (bottom panel) for galaxies in $\bm{D}_{\perp}$
bins (different line types, see key figure), for the R--type (filled
circles) and S--type (empty circles) stacked void clumps, as a
function of the normalized distance along the parallel direction. 
Vertical dotted lines indicate the positions of the two void centres.
For the case of clumps of two R--type voids, we find parallel
components of the velocity being mainly positive for
$\bm{D}_{\parallel}>0$, and negative for $\bm{D}_{\parallel}<0$. 
This implies an outward flux from the void--pair centre.

For S--type voids, is clear that the flux of halos along the parallel
direction behaves very differently than the R--type case.  The
$\bm{V}_{\parallel}$ values are negative for $\bm{D}_{\parallel}>0$
and positive for $\bm{D}_{\parallel}<0$, reflecting an inward motion
of galaxies.                       
These effects are less pronounced for the outer regions as seen in the
different line types. 
For both void types, these velocity fluxes are predominantly in the
parallel direction, since the perpendicular velocity component, shown
in the lower panel, is at most a 10 percent of the maximum parallel
velocity.
These results are in agreement with those presented in previous works
\citep{lambas_sparkling_2016, ceccarelli_sparkling_2016}, where the
dynamics of void pairs was analysed accordingly to their environments. 


\begin{figure*}
\centering
\hfill
\includegraphics[width=\textwidth,height=0.45\textwidth]{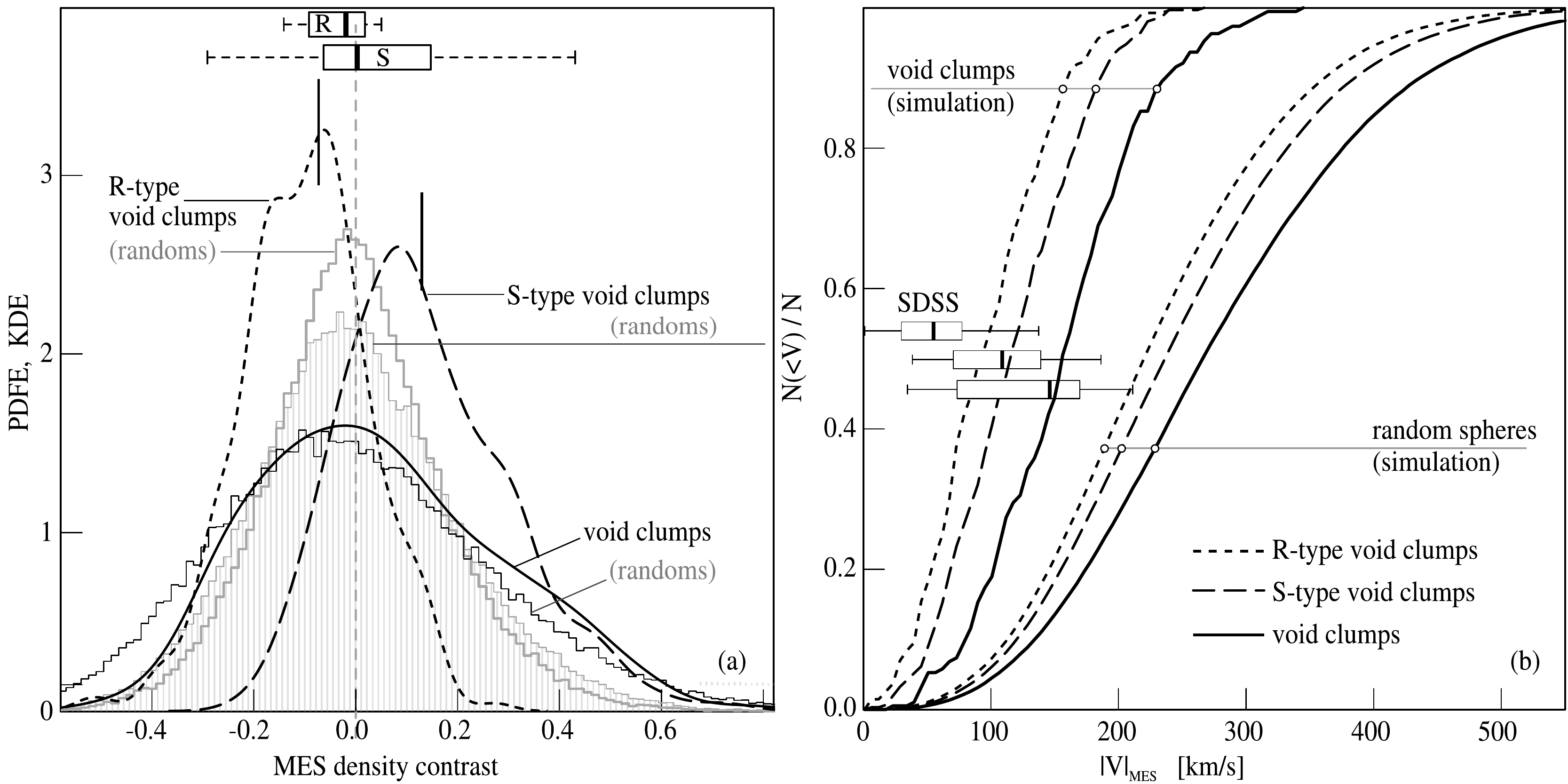}
\caption{{\it Panel (a):} Probability distribution estimates of the
   density contrast of tracers within the MESs.  Thick lines show the
   kernel density estimates of the distribution of density contrast
   values for MESs associated to all (solid), R--type (dashed) and
   S--type (dot--dashed) void clumps.  The histograms show the
   distribution estimates for randomly placed spheres with the same
   radii distribution than that of the full sample (thin black),
   R--type (grey), and S--type (grey, vertical dashed) void clumps.
   The medians of the R/S--types distributions are located with
   vertical solid lines. The box plots at the top of the panel
   correspond to R--type  and S--type void clumps in the SDSS sample.
   {\it Panel (b):} Empirical cumulative distribution functions of the
   bulk velocity magnitudes.   Curves correspond to the simulation and
   box plots to the data from SDSS.   The three curves on the left
   correspond to R--type (dotted), S--type (dashed) and the full
   sample (solid) of void clumps, respectively.   The box plots
   correspond to the same subsets in the sample of SDSS void clumps,
   according to the one curve that crosses each box.   The curves on
   the right of this panel correspond to samples of randomly placed
   spheres within the simulation box with the same radii distribution
   than those in the data with the same line type.}
\label{F_5} 
\end{figure*}

\section{Environment and void clumps motions}
\label{S_envir}

In this Section we analyse different spatial and dynamical properties
of voids in clumps, as well as the global motions of void clumps.
We also perform a similar analysis applied to the SDSS data.

\subsection{Dynamics of void pairs within clumps}

In \citet{lambas_sparkling_2016} we showed that the dynamical
behaviour of voids is characterised by a combination of two coherent
motions: approaching or receding movements between pairs of voids.
This produces a bimodal distribution of the relative velocities that
can be represented by the values of the cosines between the two
velocity vectors of each pair.
This bimodal distribution disappears once the pairs of the same type
are considered, giving rise to approaching or receding voids in the
void-in-cloud or in the void--in--void cases, respectively.
Here we explore the relation between this bimodality and the groups of
voids.
As in \citet{lambas_sparkling_2016}, we compute for each void its bulk
velocity, or simply the void velocity, by taking the average velocity
of all galaxies within a shell of $0.8$ and $1.2$ void radius. 
This is equivalent to the void bulk velocity computed in the
full radii range within the void radius
\citep{ceccarelli_sparkling_2016, lambas_sparkling_2016}. 
We then reproduce the procedure used in that work to compute the angle
$\theta$ subtended by the void clump members pairwise velocity
($\bm{\Delta V}$) and the void relative separation ($\bm{\Delta R}$).
In Fig. \ref{F_4} we show the number density of void pairs as a
function of cos($\theta$), for several selections of the void pairs
used to compute the angle.
The density distribution for the complete sample of void pairs,
separated accordingly to its void type, is estimated by computing the
histogram of the cos($\theta$) values. These distributions are shown
for reference as green shaded histograms in Fig. \ref{F_4}.

The estimation of the distributions for voids within the same clump
and for voids in different clumps are shown, for S-- and R--type
voids, with the shaded regions.
These regions are similar to the histograms, but are computed without
using bins, by fitting the empirical cumulative distribution with a
Fourier expansion, and filtering the hight frequency components that
are produced by the noise in the random sampling
\citep{berg_from_2008}.
The width of the regions indicate the resampling uncertainty computed
for each sample.
In the Figure, it can be seen that the bimodal behaviour reported by
\citet{lambas_sparkling_2016} holds for the two subsets of voids
pairs, irrespective of them being part of a clump or not, except for a
slight difference at the extreme values of cos($\theta$) in the
S--type clumps.


\subsection{Dynamics of void clumps}

In this subsection we analyse the global density contrast and the
dynamics of the large-scale regions around void clumps.
To that end, we use the Minimal Enclosing Sphere (MES) of each clump
as an approximation of how much spread is the group of voids.
The MES is defined as the smallest bounding sphere that completely
includes all the voids in the clump.
We chose to use this approach by virtue of its simplicity, which also
allows the define a centre and explore the dependence of different
properties on the distance to this centre.

We have computed the mean mass density contrast and the
total mean velocity of the galaxies inside the MESs considering
separately clumps of R--type or S--type voids.
For comparison, we have also computed these quantities in spheres
located at random positions, and with the same radii distribution than
that of the corresponding sample. 
In the left panel of Fig. \ref{F_5} we show the probability
distribution estimates of the density contrast of tracers within the
MESs. 
Thick lines show the kernel density estimates \citep[KDE,
][]{kde_r_2016} of the distribution of density contrast values for
MESs associated to all (solid), R--type (dashed) and S--type
(dot-dashed) void clumps.
The vertical marks correspond to the locations of the medians of the
R/S--type distributions. 
The histograms show the distribution estimates for randomly placed
spheres with the same distribution of radii than that of the full
sample (thin black), R--type (grey), and S--type (grey, vertical
dashed) void clumps. 
According to this figure, the MES density contrast distribution for
the complete sample of void clumps is similar to that of random
spheres with the same radii distribution.
However, when void clumps are split into R/S types, the distributions
differ significantly.
The distributions of the MES density contrast values of each type of
void clumps 
Individual R-- and S--type void clumps have narrower distributions
than the full sample and have opposite signs.
The fact that a large region containing a number of S--type voids has
a positive global density requires the presence of overdense
structures between void members of a clump to overcome their low
densities.
We notice the low number of clumps in the SDSS catalogue given the
limited its volume, which makes it difficult a direct comparison of
the distributions.
The box plots in the top of panel (a) correspond to R-- and S--type
void clumps in the SDSS sample. 
These plots are constructed using the median (central mark), the first
and third quartiles (borders of the boxes) and the extreme values in
the sample (minimum and maximum, indicated by the lines).
Although the differences are not very significant, the observations
suggest the same trend than the simulation.

In panel (b) of Fig. \ref{F_5} we show the cumulative distribution
functions of the bulk velocity moduli of the simulation void clump
MESs.
As in panel (a), the box plots correspond to the SDSS data.
The three curves on the left correspond to R--type (dotted), S--type
(dashed) and the full sample (solid) of void clumps MESs,
respectively.
The box plots are associated to the same subsets in the sample of SDSS
void clumps according to the curve that crosses each box.
The curves on the right of this panel correspond to samples of
randomly placed spheres within the simulation box with the same radii
distribution than those in the data following  the same line types.
We find results for both R-- and S--type void clumps remarkably
similar, showing less than half the bulk velocity of random spheres of
similar radii distributions.
We stress the observed similarity of R-- and S--type void clump MES
global motions in spite of their different environments and internal
dynamics.



\section{Discussion}
\label{S_discuss}

In this work we study the clustering of cosmic voids by computing
their autocorrelation function in a cosmological simulation and in a
galaxy catalogue.
Albeit the different void definitions and identification methods found
in the literature, we consider those defined as spheres that have
total densities of at most 10 per cent the mean.
We explore two different void types characterised according to the
shape of their integrated density profiles
\citep{ceccarelli_clues_2013}, following theoretical work by
\citet{sheth_hierarchy_2004}, who introduced void--in--void or
void--in--cloud scenarios.
We find significant differences in the distribution of voids with
respect to a Poisson distribution and make use of the R/S void type
definition to provide further insight on their clustering properties.
Following this line, we define clumps of voids as large regions
comprising voids of the same type according to the R/S classification.
In this work, we look forward to deepen our understanding on the
large--scale structure formation through the dynamics of voids.
The results obtained for the dynamics of voids are consistent with the
sparkling universe picture \citep{lambas_sparkling_2016,
ceccarelli_sparkling_2016}, where the large--scale structure growth
receives an imprint from void expansions and bulk motions.

\citet{padilla_spatial_2005} analysed the clustering of voids
identified in a simulation, reporting stronger clustering for the
larger voids.
This result is consistent with our findings of a higher correlation
amplitudes for R--type voids, given their larger average radii
\citep{ceccarelli_clues_2013}.
\citet{chan_large_2014} also studied the clustering of voids using
numerical simulations, although their definition of voids is different
from ours, making it difficult a direct comparison.
More recently \citet{clampitt_clustering_2016} estimated the
correlation of voids in SDSS, obtaining results consistent with our
findings.
As mentioned in previous sections, we obtain higher correlation
amplitudes which can be explained in terms of the environmental
classification.
The detection of a correlation of larger amplitudes for voids embedded
in similar large--scale environments, is a motivation to search for
clumps of voids of the same type. 
Both R-- and S--type void clumps have similar geometrical properties,
with R--type void clumps slightly more spherical.

Our environmental classification of R/S void types is a natural
scenario to study the properties and evolution of voids and their
relation to the mass distribution and dynamics. 
In \citet{paz_clues_2013}, we analysed the dynamics of void
surroundings using redshift space distortions of the void--galaxy
cross--correlations, according to the void environment, finding that
large voids are typically in an expansion phase, whereas small ones
tend to be surrounded by collapsing overdense regions.
This twofold behaviour was observed both in simulations and SDSS data.
Clustering and environment of voids have been used to analyse their
bias, providing tests for the growth of cosmic structure and measures
of cosmological parameters \citep{hamaus_probing_2015, chuang_linear_2016, hamaus_constraints_2016}.
Using redshift space distortions, \citet{achitouv_consistency_2016}
present a test to discriminate between modified gravity models.
\citet{hawken_vimos_2016} perform a similar study at larger redshifts
using VIPERS, obtaining consistent results for the measurement of the
linear growth rate.
\citet{cai_redshift_2016} also use redshift--space distortions around
voids and find that the distortion pattern depends on the type of void
being considered.
The large--scale flows of mass induced by this void evolution scenario
seem to be an essential part of structure formation, although the
effects are limited to the presence of a single void.
These previous results are in agreement with the velocity field around
void clumps defined by R/S--type voids, reported in Sec.
\ref{S_envir}, where clumps of R/S--type voids introduce
divergent/convergent large--scale flows. 
Since clumps of R/S-type voids are embedded in larger under/overdense
regions, these stream motions can be understood as driven by the clump
inner mass distribution.
The analysis of pairwise velocity of voids in clumps performed
reinforces the receding (approaching) processes dominating the
relative dynamics of R--type (S--type) voids, although we find voids
in different clumps behave similarly.
We also analyse the bulk motions of void clumps finding significantly
lower velocities than randomly placed spheres with the same radii
distribution. In the context of the sparkling universe scenario where
voids move in a coherent fashion, void clumps are dynamically
conspicuous regions.
A low bulk motion has also been reported for watershed voids
\citep{hamaus_testing_2014, sutter_life_2014}.
It must be noticed, however, that the void clumps described in this
work and the voids resulting from the ZOBOV algorithm are of a
different nature, and is not straightforward to associate both types
of regions.
On the other hand, spherical voids have been reported to have
non--negligible velocities.
This had been suggested by \citet{gottlober_structure_2003} using
spherical voids in a numerical simulation.
Also, the bulk velocities are comparable to the velocity of random
spheres of the same size \citep{ceccarelli_sparkling_2016}.
Given the different definitions and identification procedures, a
direct comparison of the results is not straightforward although their
similarity is remarkable.

This work reports on the clustering of voids, as an alternative
approach to study the large--scale distribution of mass and its
dynamics.
The prevalence of a characteristic scale for the clustering of voids
allows to define groups of nearby voids, which are suitable
laboratories to further explore the implications of the void
environments on its evolution and on the formation of the large--scale
structure of the universe.
Galaxies residing in the suburbs of void clumps display dynamical
behaviours dominated by velocities toward (away) systems defined by
S--type (R--type) of voids.
The mean velocities compare well with those derived in
\citet{paz_clues_2013} for expanding and collapsing voids and are
govern by the global density.
All these effects are consistent with the sparkling universe picture
\citep{lambas_sparkling_2016,ceccarelli_sparkling_2016}, where the
large--scale structure growth can be considered as the result of the
voids motions and evolution.

It is worth mentioning that the void statistics is limited by the low
number of voids in the observed Universe and this become more
restrictive when we consider subsamples of voids according their
global density. 
Besides the observational limitations such as low number of voids,
small volume, positions in redshift space and linearized velocities,
we obtain, when it is possible perform the comparison, compatible
results in theoretical and observational analysis. 
In the context of the new galaxy surveys such as HETDEX
\citep{hill_hetdex_2008}, Euclid \citep{euclid_2011}, SDSS-III
\citep{sdss3_2011}, VIPER \citep{2014_viper} and the Dark Energy
Survey \citep{des_2016}, the extent of the new data available hold a
promising scenario to confront and improve the results on void
clustering introduced here.

Given the relevance of void clumps on void dynamics,  studying their
relation to large structures can shed new light on large--scale flows
and the formation of the supercluster--void network.


\section*{Acknowledgments}

This work was partially supported by the Consejo Nacional de
Investigaciones Cient\'{\i}ficas y T\'ecnicas (CONICET), and the
Secretar\'{\i}a de Ciencia y Tecnolog\'{\i}a, Universidad Nacional de
C\'ordoba, Argentina. 
Plots were made using R software and post--processed with Inkscape.
This research has made use of NASA's Astrophysics Data System. 
The authors would like to thank the anonymous reviewer for their valuable
suggestions.
Funding for the SDSS and SDSS-II has been provided by the Alfred P.
Sloan Foundation, the Participating Institutions, the National Science
Foundation, the U.S. Department of Energy, the National Aeronautics
and Space Administration, the Japanese Monbukagakusho, the Max Planck
Society, and the Higher Education Funding Council for England. The
SDSS Web Site is http://www.sdss.org/.
The SDSS is managed by the Astrophysical Research Consortium for the
Participating Institutions. The Participating Institutions are the
American Museum of Natural History, Astrophysical Institute Potsdam,
University of Basel, University of Cambridge, Case Western Reserve
University, University of Chicago, Drexel University, Fermilab, the
Institute for Advanced Study, the Japan Participation Group, Johns
Hopkins University, the Joint Institute for Nuclear Astrophysics, the
Kavli Institute for Particle Astrophysics and Cosmology, the Korean
Scientist Group, the Chinese Academy of Sciences (LAMOST), Los Alamos
National Laboratory, the Max-Planck-Institute for Astronomy (MPIA),
the Max-Planck-Institute for Astrophysics (MPA), New Mexico State
University, Ohio State University, University of Pittsburgh,
University of Portsmouth, Princeton University, the United States
Naval Observatory, and the University of Washington.


\newcommand{\noop}[1]{}

\label{lastpage}

\end{document}